\documentclass[epj]{svjour}

\usepackage{latexsym}
\usepackage{graphicx}
\usepackage{fancyhdr}
\def\makeheadbox{{
 }}
\def\mytitle{My title} 
\def\myauthors{My name}  
\def\mytype{My type of session}
\def\mysession{My session}

\def\mytitle{Sneutrino-antisneutrino mixing at future colliders} 
\def\myauthors{Tuomas Honkavaara}    
\def\mytype{Contributed Talk}    
\def\mysession{Colliders - SUSY Phenomenology}

\newcommand{\bea}{\begin{eqnarray}}
\newcommand{\eea}{\end{eqnarray}}

\def\mpT{p_T \hspace{-.9em}/\;\:}

\usepackage{amssymb}

\begin{document}
\title{Sneutrino-antisneutrino mixing at future colliders}
\author{Tuomas Honkavaara\inst{1,2}
\thanks{Talk given at the 15th International Conference on Supersymmetry 
and the Unification of Fundamental Interactions, Karlsruhe, Germany, July 
26 -- August 1, 2007.}
\thanks{\emph{Email:} tuomas.honkavaara@helsinki.fi}
\and
Katri Huitu\inst{1,2}
\and
Sourov Roy\inst{3}
}                     
\institute{High Energy Physics Division, Department of Physical Sciences, 
P.O. Box 64, FIN-00014 University of Helsinki, Finland
\and 
Helsinki Institute of Physics, P.O. Box 64, FIN-00014 University of 
Helsinki, Finland
\and
Department of Theoretical Physics and Centre for Theoretical Sciences, 
Indian Association for the Cultivation of Science, Kolkata 700 032, India}
%
\date{\today}
\abstract{
\vskip -6.5cm \rightline{HIP-2007-50/TH} \vskip 6.2cm
Sneutrino-antisneutrino mixing occurs in a supersymmetric model where 
neutrinos have nonzero Majorana masses. This can lead to the sneutrino 
decaying into a final state with a ``wrong-sign charged lepton". In 
an $e^- \gamma$ collider, the signal of the associated production of an 
electron-sneutrino and the lighter chargino and their subsequent decays 
can be $e^- \gamma \rightarrow e^+ {\tilde \tau}_1^- {\tilde \tau}_1^- + 
\mpT$ where the ${\tilde \tau}_1$s are long-lived and can produce heavily 
ionizing charged tracks. This signal is free of any Standard Model 
background, and the supersymmetric backgrounds are small. 
Such a signal can be experimentally observable under certain conditions 
which are possible to obtain in an anomaly-mediated supersymmetry breaking 
scenario. Information on a particular combination of the neutrino masses 
and mixing angles can also be extracted through the observation of this 
signal. Sneutrino-antisneutrino mixing at the LHC is currently under 
study, and asymmetry considerations seem promising there.
}
%
\maketitle

\section{Introduction}

There has been a tremendous experimental progress in neutrino physics in 
recent years, and the present data from the solar and atmospheric neutrino 
experiments contain compelling evidence that neutrinos have tiny masses. 
It is widely believed that the lepton number ($L$) may be violated in 
nature and the neutrinos are Majorana particles. In this case, the 
smallness of the neutrino masses can be explained by the seesaw mechanism 
or by dimension-five nonrenormalizable operators with a generic 
structure. In the context of supersymmetric theories, such $\Delta L = 2$ 
Majorana neutrino mass terms can induce mixing between the sneutrino and  
the antisneutrino and a mass splitting ($\Delta m_{\tilde\nu}$) between 
the physical states. The effect of this mass splitting is to induce 
sneutrino-antisneutrino oscillations, and the lepton number can be tagged 
in sneutrino decays by the charge of the final state lepton. This can, for 
example, result in like-sign dilepton signals at $e^+e^-$ colliders and 
hadron colliders (see the references in~\cite{egamma}). In this talk, 
based on Ref.~\cite{egamma}, we focus on sneutrino-antisneutrino 
oscillation in the context of an $e^-\gamma$ collider.

In sneutrino-antisneutrino oscillations, which were first discussed 
in~\cite{hirschetal}, the  situation is similar to the flavour oscillation 
in the $B^0$--$\bar{B}^0$ system. Suppose the physical sneutrino states 
are denoted by $|\tilde{\nu}_1 \rangle$ and $|\tilde{\nu}_2\rangle$. An 
initially (at $t=0$) produced pure $|\tilde{\nu}\rangle$ state is related 
to the mass eigenstates as 
\bea
|\tilde{\nu} \rangle = \frac{1}{\sqrt{2}}[|{\tilde\nu}_1 \rangle + 
i|{\tilde\nu}_2 \rangle].
\eea
The state at time $t$ is 
\bea
|\tilde{\nu}(t) \rangle & = & 
\frac{1}{\sqrt{2}}[e^{-i(m_1-i\Gamma_{\tilde\nu}/2)t} |{\tilde\nu}_1 
\rangle + ie^{-i(m_2-i\Gamma_{\tilde\nu}/2)t}|{\tilde\nu}_2 \rangle], 
\nonumber \\
\eea
where the difference between the total decay widths of the two mass 
eigenstates has been neglected, and the total decay width is set to be 
equal to $\Gamma_{\tilde\nu}$. Since the sneutrinos decay, the 
probability of finding a ``wrong-sign charged lepton" in the decay of a 
sneutrino should be the time-integrated one and is given by
\bea
P(\tilde \nu \rightarrow \ell^+) = {\frac {x^2_{\tilde \nu}} 
{2(1+x^2_{\tilde \nu})}} \times B({\tilde \nu}^* \rightarrow \ell^+),
\label{eqn-prob}
\eea
where the quantity $x_{\tilde \nu}$ is defined as 
\bea
x_{\tilde \nu} \equiv \frac {\Delta m_{\tilde \nu}} {\Gamma_{\tilde \nu}},
\label{xnu}
\eea  
and $B(\tilde \nu^* \rightarrow \ell^+)$ is the branching fraction for 
$\tilde \nu^* \rightarrow \ell^+$. Here, we assume that sneutrino flavour 
oscillation is absent and the lepton flavour is conserved in the decay of 
antisneutrino/sneutrino. If $x_{\tilde \nu} \sim 1$ and if the branching 
ratio of the antisneutrino into the corresponding charged lepton final 
state is also significant, then one can have a measurable ``wrong-sign 
charged lepton" signal from the single production of a sneutrino in 
colliders.

It is evident from the above discussion that the probability of the 
sneutrino-antisneutrino oscillation depends crucially on $\Delta m_{\tilde 
\nu}$ and $\Gamma_{\tilde \nu}$. Taking into account the radiative 
corrections to the Majorana neutrino mass $m_\nu$ induced by $\Delta 
m_{\tilde \nu}$, one faces the bound~\cite{grossman-haber1} $\Delta 
m_{\tilde \nu}/m_\nu \lesssim \mathcal{O}(4\pi/\alpha)$. If we consider 
$m_\nu$ to be $\sim 0.1$~eV, then $\Delta m_{\tilde \nu} \lesssim 0.1$~keV. 
Thus, in order to get $x_{\tilde \nu} \sim 1$, one also needs the 
sneutrino decay width $\Gamma_{\tilde \nu}$ to be $\sim 0.1$~keV or so. In 
other words, this small decay width means that the sneutrino should 
have enough time to oscillate before it decays. However, such a small  
decay width is difficult to obtain in most of the scenarios widely  
discussed in the literature with the lightest neutralino ($\tilde 
\chi^0_1$) being the lightest supersymmetric particle (LSP). In this case, 
the two-body decay channels $\tilde \nu \rightarrow \nu 
\tilde\chi^0$ and/or $\tilde \nu \rightarrow \ell^- \tilde \chi^+$ 
involving the neutralinos ($\tilde \chi^0$) and the charginos ($\tilde 
\chi^+$) will open up. In order to have a decay width $\Gamma_{\tilde \nu} 
\lesssim {\mathcal O}(1)$~keV, these two-body decay modes should be 
forbidden so that the three-body decay modes $\tilde \nu \rightarrow 
\ell^- {\tilde \tau}_1^+ \nu_\tau$ and $\tilde \nu \rightarrow \nu {\tilde 
\tau}_1^\pm \tau^\mp$ are the available ones. In addition, one should get 
a reasonable branching fraction for the $\ell^- {\tilde \tau}_1^+ 
\nu_\tau$ final state in order to get the wrong-sign charged lepton 
signal. It has been pointed out in Ref.~\cite{grossman-haber1} that, in 
order to achieve these requirements, one should have a spectrum 
\bea
m_{{\tilde \tau}^\pm_1} < m_{\tilde \nu} < m_{\tilde \chi^0_1}, m_{\tilde 
\chi^\pm_1}, 
\label{spectrum}
\eea   
where the lighter stau (${\tilde \tau}^\pm_1$) is the LSP. However, having 
${\tilde \tau}^\pm_1$ as a stable charged particle is strongly disfavoured 
by astrophysical grounds. This could be a\-void\-ed, for example, 
by assuming a very small $R$-par\-i\-ty-vi\-o\-lat\-ing coupling which 
induces the decay ${\tilde \tau}_1 \rightarrow \ell \nu$ but still allows 
${\tilde\tau}_1$ to have a large enough decay length to produce a heavily 
ionizing charged track inside the detector. The spectrum (\ref{spectrum}) 
can be obtained in some part of the parameter space in the context of 
anomaly-mediated supersymmetry breaking (AMSB)~\cite{amsb1} with $\Delta 
m_{\tilde \nu} \lesssim \mathcal{O}(4\pi m_\nu/\alpha)$. Hence, AMSB seems 
to have a very good potential to produce signals of 
sneu\-trino-antisneutrino oscillation which can be tested in colliders.

The minimal AMSB (mAMSB) model is described by the following parameters: 
the gravitino mass $m_{3/2}$, the common scalar mass parameter $m_0$, the 
ratio of Higgs vacuum expectation values $\tan\beta$, and sgn($\mu$). 

In the mAMSB, assuming that the small neutrino masses are generated by
high energy supersymmetry preserving dynamics at an energy scale $M$ far 
above the weak scale, the relevant part of the superpotential and the soft 
SUSY breaking interactions are given by~\cite{like-sign-norp}
\bea
\Delta W_{\rm eff} = \frac {\Phi_0}{M}\lambda_{ij}(L_iH_2)(L_jH_2),
\label{eqndelw}
\eea 
\bea
\Delta{\cal L}_{\rm soft} = \frac {C_{ij}\lambda_{ij}}{M}({\tilde \ell}_i 
h_2) 
({\tilde \ell}_j h_2),
\label{eqndelsoft}
\eea
where $H_2$ is the Higgs doublet superfield giving masses to the up-type 
quarks and $L_i$ are the lepton doublet superfields, with the scalar 
components $h_2$ and ${\tilde \ell}_i$, respectively. $\Phi_0$ is the Weyl 
compensator superfield, $C_{ij} \approx M_{\mathrm{aux}}$, and $\lambda$ 
is a matrix in flavour space. After the electroweak symmetry breaking, 
$\Delta W_{\rm eff}$ gives a neutrino mass matrix $(m_\nu)_{ij} 
= {\frac {2}{M}} \lambda_{ij}{\langle h_2 \rangle}^2$. Including the 
$\Delta L = 2$ contribution from Eq. (\ref{eqndelsoft}) to the sneutrino 
mass-squared matrix in the AMSB scenario, one obtains the sneutrino mass 
splitting~\cite{like-sign-norp}
\bea
(\Delta m_{\tilde \nu})_{ij} \approx \frac{M_{\mathrm{aux}}}{m_{\tilde 
\nu}} (m_\nu)_{ij} = \mathcal{O}(4\pi(m_\nu)_{ij}/\alpha). 
\label{eqn-deltamsnu} 
\eea

At the $e^-\gamma$ collider, we are interested in $(\Delta m_{\tilde 
\nu})_{ee} = \frac{4\pi}{\alpha}(m_\nu)_{ee}$, since we want to produce an 
electron-sneu\-trino. The one-loop contribution to the neutrino mass 
coming from the sneutrino mass splitting can be 
significant~\cite{grossman-haber1}. Writing this total contribution as 
$(m_\nu)_{ee} = (m_\nu)^0_{ee} + (m_\nu)^1_{ee}$, we use the constraint 
$|(m_\nu)_{ee}| < 0.2$~eV coming from the searches for the neutrinoless 
double beta decay. Here, $(m_\nu)^0_{ee}$ is the tree-level value 
discussed in Eq. (\ref{eqn-deltamsnu}) and $(m_\nu)^1_{ee}$ is the 
one-loop contribution.

The way to obtain very high energy photon beams is to induce laser 
back-scattering off an energetic $e^\pm$ beam~\cite{egammacollider}. The 
use of perfectly polarized electron and photon beams maximizes the signal 
cross section, although, in reality, it is almost impossible to achieve 
perfect polarizations. For the laser beam, perfect polarization is 
relatively easy to obtain, and we shall use $|P_L| = 1$. However, the same 
is not true for electrons or positrons, and  we use $|P_b| = |P_{e^-}| = 
0.8$ as a conservative choice. Since we want to produce the sneutrino in 
this study, the $e^-$ should be left-polarized, i.e. $P_{e^-} = -0.8$. In 
order to  improve the monochromaticity of the outgoing photons, the laser 
and the $e^\pm$ beam should be oppositely polarized, which means $P_L 
\times P_b < 0$.

\section{Signal and backgrounds}

We are interested in the production process $e^- \gamma \to {\tilde \nu}_e 
{\tilde \chi^-_1}$ and then look at the oscillation of the ${\tilde 
\nu}_e$ into a ${\tilde \nu}^*_e$. The resulting antisneutrino then decays 
through the three-body channel ${\tilde \nu}^*_e \to e^+ {\tilde \tau}_1^- 
{\bar \nu}_\tau$ with a large branching ratio. The chargino $\tilde 
\chi^-_1$ subsequently decays into a ${\tilde \tau}_1^-$ and an 
antineutrino ($\bar{\nu}_\tau$). The neutrinos escape detection and give 
rise to an imbalance in momentum. The signal is then
\bea
e^- \gamma \to {\tilde \nu}_e {\tilde \chi^-_1} \to e^+ + {\tilde 
\tau}_1^- + {\tilde \tau}_1^- + \mpT,
\label{signal}
\eea   
where the two ${\tilde \tau}_1^-$s are long-lived and can produce heavily 
ionizing charged tracks inside the detector after traversing a macroscopic 
distance. The positron serves as the trigger for the event. We assume that 
the ${\tilde \tau}_1^-$ decays through a tiny $R$-parity-violating 
coupling $\lambda_{233} = 5 \times 10^{-9}$ into charged lepton + neutrino 
pairs so that a substantial number of events do have a reasonably large 
decay lengths for which the displaced vertex may be visible.

We select the signal events in Eq. (\ref{signal}) according to the 
following criteria:
\begin{enumerate}
\item The transverse momentum of the positron must be large enough:  
$p^{e^+}_T> 10$~GeV.
\item The transverse momentum of the ${\tilde \tau}_1^-$s must satisfy 
$p^{{\tilde \tau}_1}_T> 10$~GeV.
\item The positron and both the staus must be relatively  central, i.e. 
their pseudorapidities must fall in the range $|\eta^{{e^+},{\tilde 
\tau}_1}|< 2.5$.
\item The positron and the staus must be well-separated from each other: 
i.e. the isolation variable $\Delta R \equiv \sqrt{(\Delta \eta)^2 + 
(\Delta \phi)^2}$ (where $\eta$ and $\phi$ denote the separation in 
rapidity and the azimuthal angle, respectively) should satisfy $\Delta R > 
0.4$ for each combination.
\item The missing transverse momentum $\mpT > 10$~GeV.
\item Both the heavily ionizing charged tracks due to the long-lived staus 
should have a length $\ge 5$~cm.
\end{enumerate}

The signal (\ref{signal}) is free of any Standard Model (SM) backgrounds 
when the $\lambda_{233}$ coupling is small. However, there are backgrounds 
from SUSY processes~\cite{egamma} of which $e^- \gamma \to \tilde{e}_L^- 
{\tilde \chi^0_1} \to e^+ + \tilde{\tau}_1^- + \tilde{\tau}_1^- + \mpT$ is 
the most important one. If the $\lambda_{233}$ coupling is larger, then 
the staus decay more rapidly, in which case there are SM backgrounds 
coming from $e^- \gamma \to W^-W^- W^+ \nu_e \to  e^+ \ell^- \ell^- + 
\mpT$ where $\ell =\mu,\tau$. However, these backgrounds are quite 
small~\cite{pilaftsis}.

\begin{figure}[t]
\vspace*{-2.7in}
\hspace*{-2.3cm}
\includegraphics[width=0.9\textwidth]{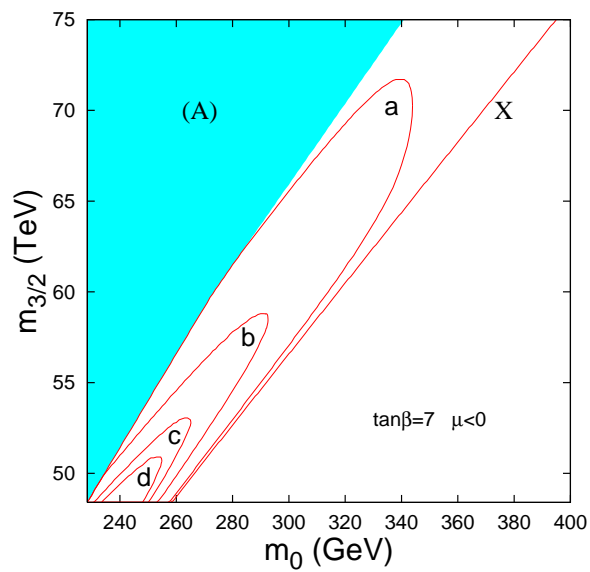}
\caption{Parameter regions with $\tan\beta=7$ and $\mu<0$. The area (A)
represents the parameter region forbidden by the stau mass bound. The 
mass spectrum (\ref{spectrum}) is obtained in the region between the area 
(A) and the line X. Assuming an integrated luminosity of $500$~fb$^{-1}$ 
at $\sqrt{s_{ee}}=500$~GeV, the numbers of positron events per year inside 
the contours are: (a) $N_e \geq 50$, (b) $N_e \geq 500$, (c) $N_e \geq 
1000$ and (d) $N_e \geq 1300$ for $(m_\nu)^0_{ee} = 0.079$~eV so that the 
total contribution $(m_\nu)_{ee} \approx 0.2$~eV, while satisfying $N_e 
\geq 5 \sqrt{N_e+N_\mathrm{B}}$.}
\label{param500}       
\end{figure}

\begin{figure}[t]
\vspace*{-2.75in}
\hspace*{-2.3cm}
\includegraphics[width=0.9\textwidth]{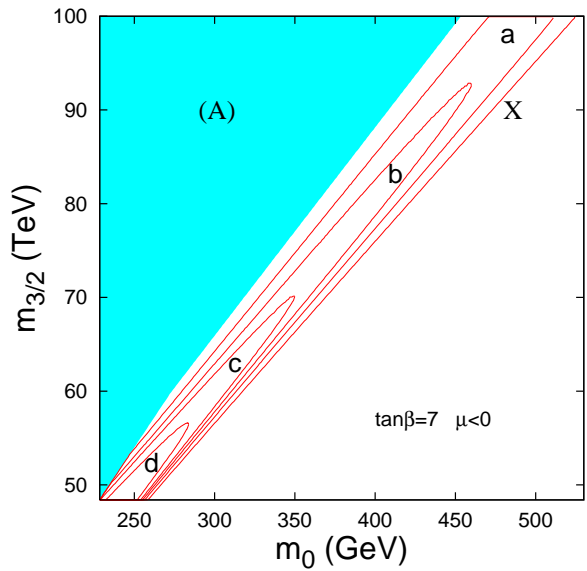}
\caption{Parameter regions with $\tan\beta=7$ and $\mu<0$. The area (A) 
represents the parameter region forbidden by the stau mass bound. The mass 
spectrum (\ref{spectrum}) is obtained in the region between the area (A) 
and the line X. Assuming an integrated luminosity of $500$~fb$^{-1}$ at 
$\sqrt{s_{ee}}=1$~TeV, the numbers of positron events per year inside the 
contours are: (a) $N_e \geq 100$, (b) $N_e \geq 200$, (c) $N_e \geq 300$ 
and (d) $N_e \geq 500$ for $(m_\nu)^0_{ee} = 0.079$~eV so that the total 
contribution $(m_\nu)_{ee} \approx$ $0.2$~eV, while satisfying $N_e \geq 5 
\sqrt{N_e+N_\mathrm{B}}$.}
\label{param1000}       
\end{figure}

In Fig.~\ref{param500}, we show our results for the total number of 
positron events for a machine operating at $\sqrt{s_{ee}} = 500$~GeV with 
$500$~fb$^{-1}$ integrated luminosity after imposing the kinematical cuts 
discussed above, while satisfying $N_e \geq 5 \sqrt{N_e+N_\mathrm{B}}$, 
where $N_e$ is the number of signal events and $N_B$ is the number of 
background events. The region marked by (A) corresponds to a lighter stau 
mass of less than $86$~GeV (see~\cite{egamma} for the references for 
different experimental constraints). The area below the line X does not 
satisfy the mass hierarchy of Eq.~(\ref{spectrum}). Thus, the allowed 
region in the ($m_0$--$m_{3/2}$) plane is the one between the area (A) and 
the line X. The other experimental constraints which we have used are the 
mass of the lighter chargino ($m_{\tilde \chi^\pm_1} > 104$~GeV), the mass 
of the sneutrino ($m_{\tilde \nu} > 94$~GeV) and the mass of the lightest 
Higgs boson ($m_h > 113$~GeV). In Fig. \ref{param1000}, we show a similar 
plot in the ($m_0$--$m_{3/2}$) plane for a machine operating at 
$\sqrt{s_{ee}} = 1$~TeV with other inputs remaining the same.

\begin{figure}[t]
\includegraphics[width=0.65\textwidth]{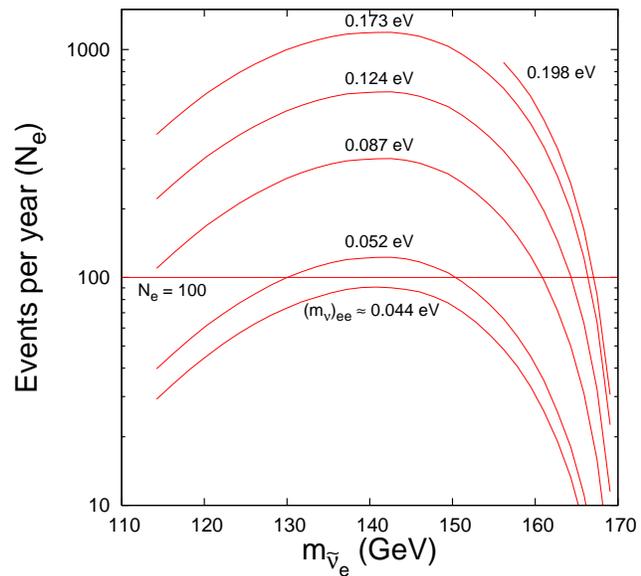}
\caption{Number of events ($N_e$) per year (with an integrated luminosity 
of $500$~$\mathrm{fb}^{-1}$) as a function of $m_{{\tilde\nu}_e}$ for 
different choices of $(m_\nu)^0_{ee}$ as dicussed in the text. Here, 
$\tan\beta=7$, $\mu<0$ and $m_{3/2} = 50$~TeV. The values of the total 
contribution $(m_\nu)_{ee}$ corresponding to each line are shown in the 
figure. The horizontal line stands for $N_e = 100$ satisfying $N_e \geq 
5\sqrt{N_e+N_\mathrm{B}}$.}
\label{fig-mnu} 
\end{figure}

Let us then discuss the change in the number of events when $(m_\nu)^0_{ee}$ is
varied in such a way that it is consistent with the upper limit of 
$0.2$~eV for the total contribution $(m_\nu)_{ee}$. For this, we choose a 
machine operating at $\sqrt{s_{ee}} = 500$~GeV. Evidently, larger values 
of $(m_\nu)^0_{ee}$ give a larger cross section. This is also shown in 
Fig. \ref{fig-mnu} for a sample choice of $m_{3/2} = 50$~TeV, $\tan\beta = 
7$ and $\mu < 0$. Assuming an integrated luminosity of 
$500$~$\mathrm{fb}^{-1}$, we have plotted the number of events per year as 
a function of $m_{{\tilde \nu}_e}$ for different choices of 
$(m_\nu)^0_{ee}$. The curves from below correspond to $(m_\nu)^0_{ee} = 
0.018$~eV, $0.021$~eV, $0.035$~eV, $0.05$~eV, $0.07$~eV and $0.081$~eV. 
The corresponding values of the total contribution $(m_\nu)_{ee}$ are 
shown in the figure. The horizontal line gives $N_e = 100$ per year. This 
figure tells us that if we demand the value of $N_e$ to be $\geq 100$, so 
that the signal significance is $\geq 5\sigma$, then we can probe the 
value of $(m_\nu)_{ee}$ down to $\approx 0.05$~eV. On the other hand, the 
current upper limit of $0.2$~eV on $(m_\nu)_{ee}$ sets the upper limit of 
$(m_\nu)^0_{ee} \approx 0.081$~eV. The topmost curve in this figure starts 
from a slightly higher value of $m_{{\tilde \nu}_e}$, since the bound on 
$(m_\nu)_{ee}$ is not satisfied before that. This figure can also be used to 
extract the value of $(m_\nu)_{ee}$ with the knowledge of the number of events 
and other masses.

\section{Summary and discussion}

We have discussed the potential of an electron-photon collider to 
investigate the signature of $\tilde\nu_e$--${\tilde\nu}_e^*$ mixing in an 
AMSB model which can accommodate $\Delta L = 2$ Majorana neutrino masses. 
A very interesting feature of such models is that the 
sneutrino-antisneutrino mass splitting $\Delta m_{\tilde\nu}$ is naturally 
large and is $\mathcal{O}(4\pi m_\nu/\alpha)$. On the other hand, the 
total decay width of the sneutrino is sufficiently small in a significant 
region of the allowed parameter space of the model. These two features 
enhance the possibility of observing sneutrino oscillation signal in 
various colliders. We have demonstrated that the associated production of 
the lighter chargino and the sneutrino at an $e^-\gamma$ collider could 
provide a very clean signature of such a scenario. In addition, this 
signal can be used to determine $(m_\nu)_{ee}$ which provides important 
information on a particular combination of the neutrino masses and mixing 
angles which is not possible to obtain from neutrino oscillation 
experiments.

Sneutrino-antisneutrino mixing can also be probed in $pp$ collisions at 
the LHC. The asymmetry between various cross sections can probably 
indicate whether there is sneutrino oscillation or not. This study is in 
progress, and the asymmetry considerations seem prom\-is\-ing~\cite{lhc}.

\section*{Acknowledgements}
This work was supported by the Academy of Finland (Project number 115032).

\end{document}